\documentclass[a4paper]{article}
\usepackage{INTERSPEECH2021}

\title{Real-time Speaker counting in a cocktail party scenario using Attention-guided Convolutional Neural Network}
\name{Midia Yousefi, John H.L. Hansen}
\address{
  Center for Robust Speech Systems (CRSS), Erik Jonsson School of Engineering,\\
University of Texas at Dallas, Richardson, Texas, USA}
\email{midia.yousefi@utdallas.edu, john.hansen@utdallas.edu}

\begin{document}

\maketitle

\begin{abstract}

Most current speech technology systems are designed to operate well even in the presence of multiple active speakers. However, most solutions assume that the number of co-current speakers is known. Unfortunately, this information might not always be available in real-world applications. In this study, we propose a real-time, single-channel attention-guided Convolutional Neural Network (CNN) to estimate the number of active speakers in overlapping speech. The proposed system extracts higher-level information from the speech spectral content using a CNN model. Next, the attention mechanism summarizes the extracted information into a compact feature vector without losing critical information. Finally, the active speakers are classified using a fully connected network. Experiments on simulated overlapping speech using WSJ corpus show that the attention solution is shown to improve the performance  by almost 3\%  absolute over conventional temporal average pooling. The proposed Attention-guided CNN achieves 76.15\% for both Weighted Accuracy and average Recall, and 75.80\% Precision on speech segments as short as 20 frames (i.e., 200 ms).  All the classification metrics exceed 92\% for the attention-guided model in offline scenarios where the input signal is more than 100 frames long (i.e., 1s).  
\end{abstract}


\section{Introduction}
\label{sec:intro}

Spontaneous conversations often contain segments of overlapping speech (i.e., segments with multiple active speakers). This phenomena known as cocktail party \cite{cherry1953some,haykin2005cocktail, yousefi2018assessing}, has been studied for several decades \cite{bronkhorst2000cocktail,yousefi2020block, yousefi2016supervised}, yet still represents a major source of error in many speech technology applications such as Speaker Identification \cite{boakye2008overlapped,qian2018past,yousefi2020frame} and Automatic Speech Recognition \cite{manohar2019acoustic}. Many researcher developed systems that can trace and recognize multiple speakers \cite{qian2018past}. However, the majority of these systems assume that the number of concurrent sources is known in advance \cite{yu2017permutation,luo2018tasnet, yousefi2019probabilistic}, which is not a realistic assumption in most applications. Therefore, a real-time, simple, yet effective speaker count estimation is needed to bridge the gap between research and real-world applications \cite{von2019all}.

The majority of developed solutions for estimating the maximum number of active speakers mimic the process of human perception. Since humans use spatial information based on the acoustic signals they receive from their ears, many researchers calculate the Direction of Arrival (DoA) derived from multi-channel speech signals to count and localize the  
number of active sources. However, this approach is not practical in scenarios where only a single-channel speech recording is available. In 2003, Arai \cite{arai2003estimating} proposed the first single-channel speaker counting system based on modulation characteristics of speech. He established the estimation algorithm by defining the curve of ``equivalent number of speakers" derived from the region of  modulation  frequency between 2 and 8 Hz. He observed that when a speaker is talking, the speech  signal typically has a distinct modulation spectrum with a peak around 4-5 Hz. However, as the number of simultaneous speakers increases, the modulation pattern becomes more complex. This approach was not efficient as it relied on very precise thresholds for an  accurate estimation of simultaneous speakers. 
The authors in \cite{xu2013crowd++} introduced an unsupervised machine learning approach known as Crowd++, which used agglomerative hierarchical clustering to estimate the number of speakers in naturalistic audio. In a more recent study \cite{stoter2018countnet}, CountNet was proposed which used a supervised machine learning method for speaker estimation. In 2019, Andrei \cite{andrei2019overlapped} trained several Convolutional Neural Network architectures to count the number of active speakers in speech segments ranging from 100 to 500 frames. The aforementioned studies demonstrate the capability of Deep Neural Networks in addressing co-current speaker counting, however, none are capable of being deployed in real-time applications as they depend on at at least 2s of speech for an accurate estimation.

In this study, we propose an attention-guided architecture to count the number of co-current talkers on segments as short as  0.2 sec suitable for real-time applications. The remainder of the paper is organized as follows. In Sec \ref{sec:prob}, the problem is outlined. Sec \ref{sec:sys} describes the proposed method in detail. Sec \ref{sec:exp} presents results, finally the conclusions are discussed in Sec \ref{sec:con}. 

\section{Problem setup}
\label{sec:prob}

\begin{figure}
\centering
\includegraphics[width=9cm]{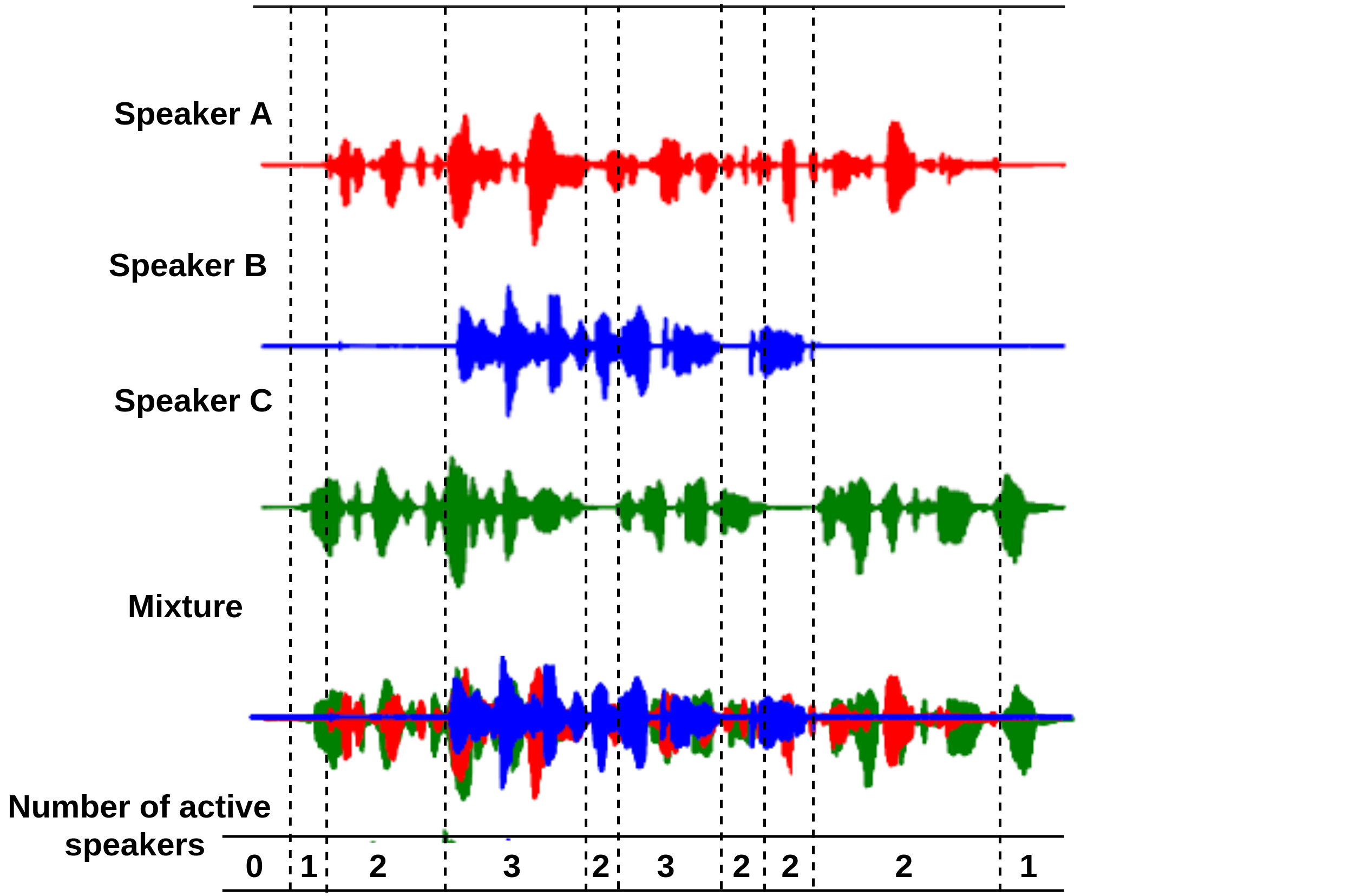}
\caption{Data generation procedure. }
\label{fig:data}
\vspace{-0.5cm}
\end{figure}

Most previous studies \cite{stoter2018countnet,andrei2019overlapped} have employed simulated overlapping speech for speaker counting because naturalistic data, such as AMI \cite{carletta2005ami}, contains only 5-10\% of overlap speech \cite{von2019all}, which is not sufficient for training DNNs. For conducting experiments, we generate a simulated dataset consisting of different scenarios of multiple active speakers using the same procedure as \cite{yu2017permutation,luo2018tasnet}. The simulated dataset is based on WSJ corpus \cite{garofolo1993csr} in which random utterances from random speakers are summed together with a Signal to Interference Ratio (SIR) chosen randomly between 0-5 dB using a uniform distribution. We generate a working corpus of 2-speaker and 3-speaker mixture speech utterances. However, as depicted in Fig \ref{fig:data}, labeling is accomplished by considering speaker activity within each utterance. Since performance of a the Speech Activity Detector is crucial in counting the active speaker within short segments, we integrate speech activity detector into speaker-count estimation by adding a non-speech class at the output. For this purpose, we add non-speech utterances to the simulated dataset, which are either silence or environmental noise. Additionally, single-speaker utterances are also added to the corpus to balance the final dataset. Based on the generated dataset, our task is a 4-class classification problem, where each segment is classified as either non-speech, 1-speaker, 2-speaker, or 3-speaker segment. The details of the simulated corpus is summarized in Table \ref{tab:data}.

\section{System design}
\label{sec:sys}

The speaker-count estimation is generally performed in three steps: \emph{(i)} extracting higher-level features from speech that are related to the number of speakers, \emph{(ii)} summarizing the frame-level extracted features into an utterance-level feature vector, and \emph{(iii)} classifying the number of speakers in the feature space. Additionally, since one of the output classes is non-speech, the proposed system can also be considered as a combined real-time Speech Activity Detector (SAD) that detects non-speech segments as short as 0.2 seconds.

\begin{table}[b]
\centering

 \begin{tabular}{||c c c c||}
 \hline
 Class & Train & Cross validation & Test \\ [0.5ex] 
 \hline\hline
 Non-speech & 5000 & 500 & 500 \\ 
 1-speaker & 5000 & 500 & 500 \\
 2-speaker & 5000 & 500 & 500 \\
 3-speaker & 5000 & 500 & 500 \\
 \hline
 total & 20000 & 2000 & 2000 \\ 
 \hline
 \end{tabular}
 
 \caption{Number of the utterances in each class.}
\label{tab:data}
\end{table}

\begin{figure}
\centering
\includegraphics[width=8.8cm, height=4.5cm]{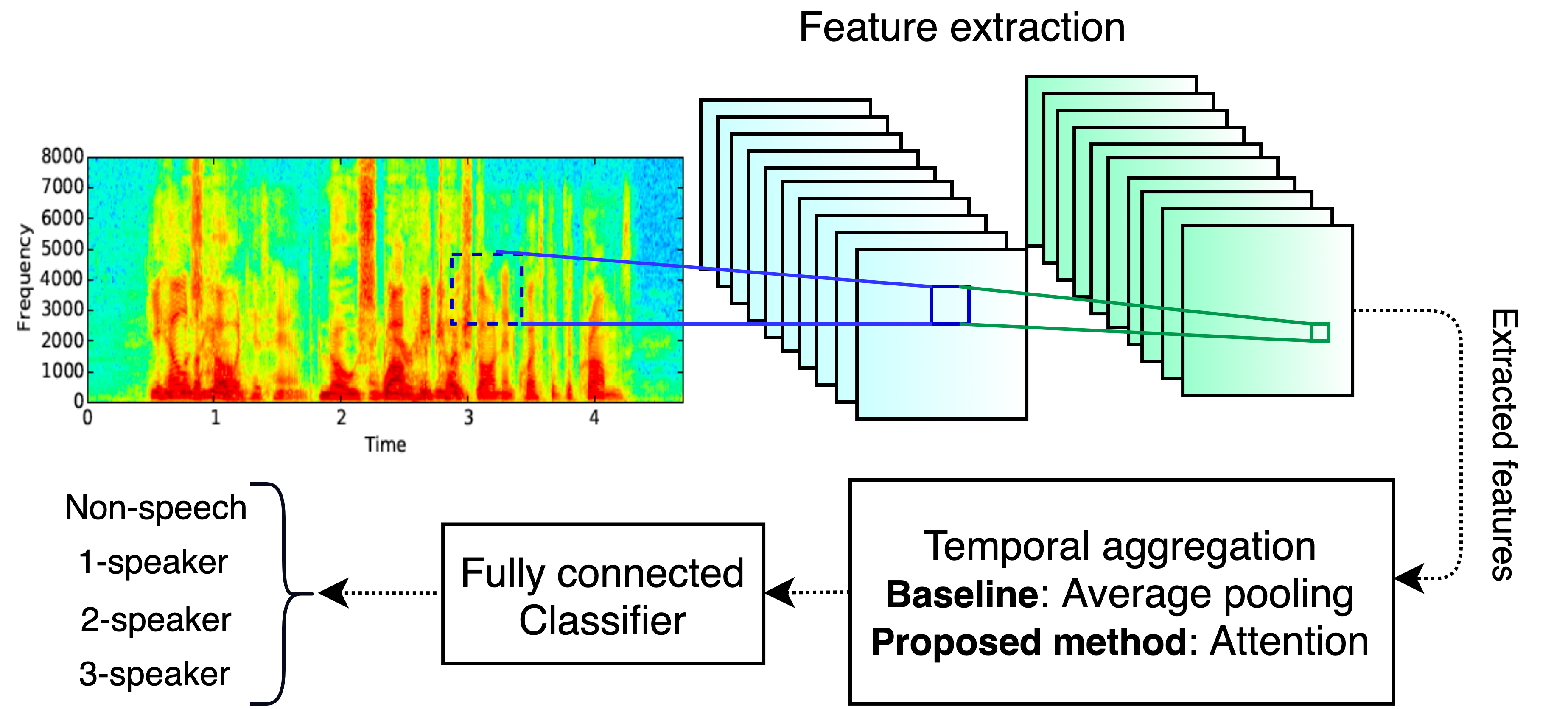}
\caption{The proposed speaker counting architecture. }
\label{fig:model}
\end{figure}

The proposed architecture is shown in Fig. \ref{fig:model}, where the 2D Convolutional Neural Network (CNN) is used for higher-level feature extraction. Next, the extracted feature map, which is a $K\times M$ matrix with a feature dimension  $K$  and the time step $M$, is passed  to the  Temporal aggregation block to be summarized into a feature vector of size $K\times 1$. This block condenses the information in the feature map across the time dimension. Finally, a fully connected (FC) network is employed to classify the feature vectors in the embedding space into one of 4 classes.

The most important block in the proposed architecture is the temporal aggregation step where the feature map should be summarized without losing important information. A typical time aggregation method is to use average pooling over time dimension which has been used most often in previous studies \cite{stoter2018countnet,andrei2019overlapped, wang2020speaker}. However, if one speaker is active for only a small portion of the segment, then using temporal averaging may suppress that speaker from the feature map, therefore resulting in a false estimation of the number of active speakers. 

For addressing this challenge, we incorporate an attention mechanism into the system to perform the temporal aggregation step. The main concept of attention is to compress all necessary information of an input sequence into a fixed-length vector \cite{vaswani2017attention}. To achieve this, the attention mechanism first finds the most informative regions in the extracted feature map, then assigns reasonable weights to those regions \cite{bahdanau2014neural}. Therefore, the model focuses on regions with important elements related to the number of active speakers \cite{galassiattention}.  In order to find relevant information and calculate their corresponding weights, the input feature map is transformed into two alternate embedding spaces called (key,value) using two fully connected layers \cite{vaswani2017attention}:

\begin{align}
K = W_{k}\times X_{feat} \\
V = W_{v} \times X_{feat}
\label{eq:k-v}
\end{align}
where, $K$ and $V$ stand for key and value respectively. $W_{k}$ and $W_{v}$ are fully connected layers that perform a linear transformation of the extracted feature map $X_{feat}$. The attention mechanism is a fully connected layer with a trainable input vector known as query $q$. Attention finds the most informative regions in the extracted feature map by comparing the key of each time step with the query using dot product \cite{vaswani2017attention}:

\begin{equation}
R = \dfrac{qK}{\sqrt{d_{k}}}
\label{eq:att1}
\end{equation}
where $d_k$ is the key dimension and $R$ is the similarity level of the key and the trained query. Higher values of $R$ reflects the similarity of the trained query with the key value of that specific region in the feature map.
Once the similarity metric $R$ is derived, a weight for each region is generated by applying a Softmax activation function to the similarity metric $R$. Finally, the value of the feature map is multiplied by the attention weights as \cite{vaswani2017attention}:

\begin{equation}
\hat{X}_{feat} = \text{softmax}(R)V^{T}.
\label{eq:att}
\end{equation}
Here, the $\hat{X}_{feat}$ is the weighted average  of the extracted feature map $X_{feat}$, for which the weights are calculated by the attention mechanism. Thus, the information in the extracted feature map is summarized into a feature vector with dimension $M \times 1$. Attention mechanism is a superior technique compared to average pooling over a time dimension because it assigns higher weights to more informative regions. After performing the temporal aggregation using attention, the feature vector calculated at the output of attention is passed to the FC classifier followed by a Softmax activation function for the final decision.

\section{Experiments, Results, and discussion}
\label{sec:exp}

\textbf{Dataset} -- the generated dataset explained in Sec \ref{sec:prob} is used for evaluating performance of speaker count estimation. Although higher-level feature extraction is performed in the proposed system shown in Fig \ref{fig:model}, we do not use the raw waveform for training the network. The reason behind this is that the time domain waveforms are dense and using them directly for network training is not computationally efficient. Thus, we extract Log Mel FilterBank (LMFB) features for network training. In order to extract LMFB features, we first apply a pre-emphasis filter $y(t) = x(t) - 0.97x(t-1) $ to boost the high frequencies. Next, we calculate 512-dim magnitude spectra computed over a frame size of 25 ms with 10 ms frame shift with 16kHz sampling frequency. The energy of each frame is also derived. A set of 40 triangular filterbanks are applied on the energy of the frames. The logarithm output is the LMFB features. 

\begin{figure}
\centering
\includegraphics[width=7.5cm]{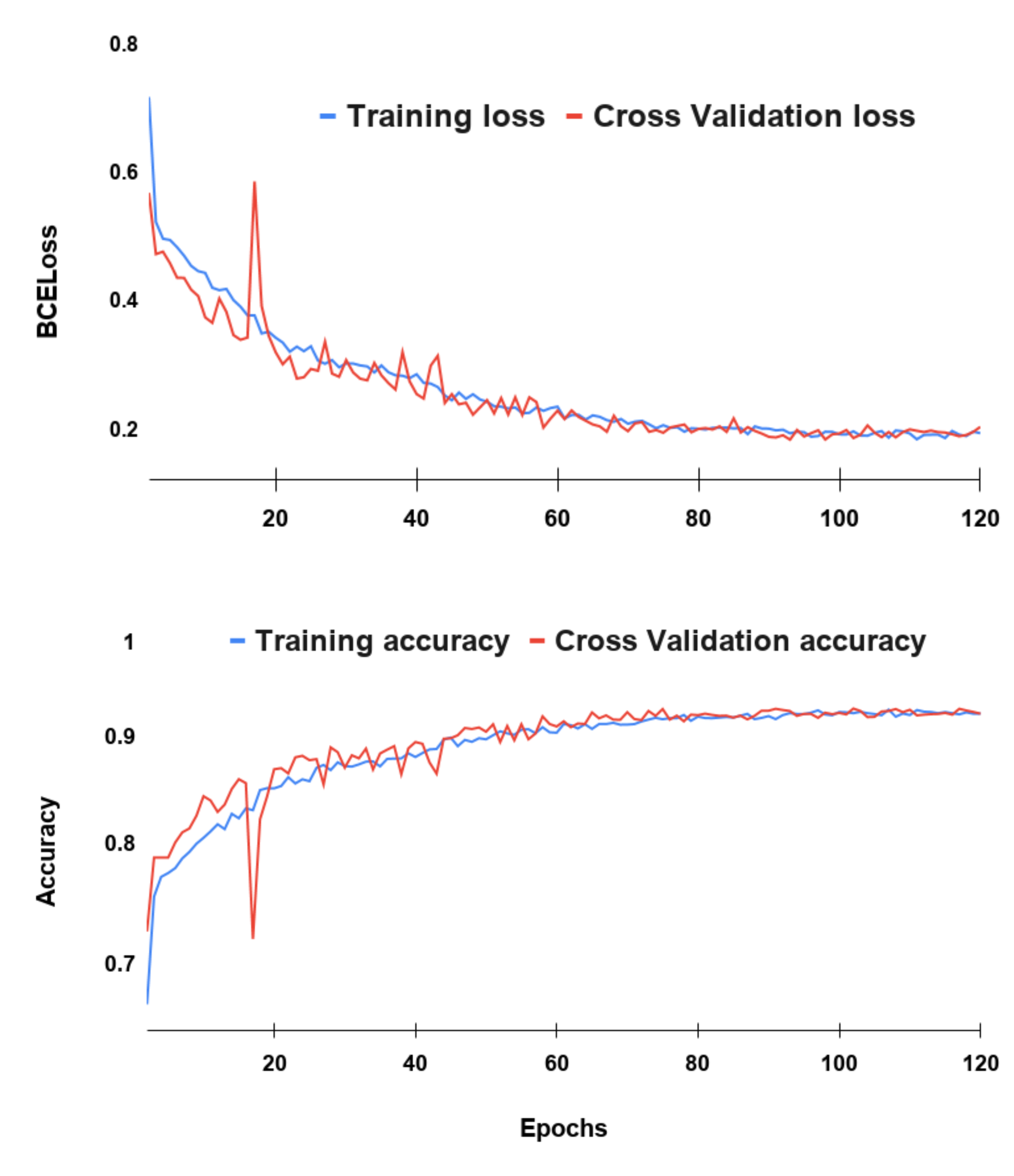}
\vspace{-0.4cm}
\caption{The training behavior of the proposed architecture. }
\vspace{-0.4cm}
\label{fig:loss}
\end{figure}

\textbf{Evaluation metrics} -- for evaluating the effectiveness of the proposed system, we use evaluation metrics based on the confusion matrix using Accuracy, Weighted Accuracy,  Precision,  Recall,  and F-score. Accuracy is the ratio between the number of correct predictions divided by the total number of speech segments. Weighted Accuracy is Accuracy calculated within each class and it is equal to the average Recall. Precision expresses the ratio of the correct predictions in each class to the total number of predicted segments for that class. Conversely, Recall is measured as the ratio of correct predictions to the total number of segments within each class. F-score is another useful measure defined as the harmonic mean of Recall and Precision.

\textbf{Model} -- the  hyper-parameters of the network are learned using the cross validation set. The choice of 8 2D convolutional layers with 128 output channels, kernel size of 5*5,  and 2 FC layers with 256 neurons each is optimum. The ReLU activation function is used between all layers. Also, the layers are initialized using Kaiming initialization \cite{he2015delving}. The output of the classifier is passed to a Softmax activation function for the final decision. The  network parameters  are  updated  by  the  gradients  of Negative Log Likelihood Loss using a Stochastic Gradient Descent (SGD) optimizer with an initial learning rate tuned to 0.01. The training process is completed by performing the early stopping \cite{zhang2016understanding}. The maximum number of epochs is set to 500, batch size set to 128, and the learning rate is decreased by a factor of 0.7 if the cross validation loss improvement is less than 0.001 for two successive epochs. The early stopping is performed if no improvement is observed on the validation set once the learning rate is decayed six times. The Training and cross validation loss and Accuracy of the network with the selected hyper-parameters are shown in Fig \ref{fig:loss}, which depicts the ability of the network to generalize to the unseen data samples in the development phase. Also, the scheduler which performs the early stopping terminates the training session after completing 120 epochs where the model convergences to the optimum local minima.

\textbf{Results and discussion} -- as mentioned in Sec \ref{sec:sys}, in the proposed method, we focus on the temporal aggregation block, which takes the extracted feature map and summarizes its information into a feature vector. The motivation behind this is that we estimate the number of classes on speech segments with a duration range of 0.2-1 sec.  Since, each segment has only one label, the temporal dimension of the feature map must be summarized into a single vector. As a baseline, the conventional temporal Average pooling is performed to squeeze the feature map into a feature vector. The results of the experiments with Average pooling is presented in Table \ref{tab:res}. Experiments are performed on speech segments with 20 and 100 frame duration (i.e, 200ms to 1s). The baseline achieves  73.85 \% Weighted Accuracy which is the same as average Recall, and 73.18\% average precision over 4 classes for 20-frame input speech. The F-score is 73.51\% for the real-time classification. However, if the speech context increases by adding more frames, the classification performance improves by a great extent. For 100-frame speech segments, the baseline  performance exceeds 90\% for all classification metrics. 

\begin{table}[t]

\begin{minipage}{.5\textwidth}
\centering
\begin{tabular}{@{\extracolsep{1pt}}lcccc} 
\\\hline 
\hline \\
20 frames & \multicolumn{1}{c}{Accuracy} & \multicolumn{1}{c}{Precision} & \multicolumn{1}{c}{Recall} & \multicolumn{1}{c}{F-score} \\ 
\hline \\ 
Average pool. & 73.85\% & 73.18\% & 73.85\% & 73.51\%\\ 
Attention & \textbf{76.15\%} & \textbf{75.80\%} & \textbf{76.15\%}  & \textbf{75.97\%}\\
\hline \\
\end{tabular} 
\end{minipage}%
\vspace{-0.2cm}
\begin{minipage}{.5\textwidth}
\centering
\begin{tabular}{@{\extracolsep{1pt}}lcccc} 
\\\hline 
\hline \\
100 frames & \multicolumn{1}{c}{Accuracy} & \multicolumn{1}{c}{Precision} & \multicolumn{1}{c}{Recall} & \multicolumn{1}{c}{F-score} \\ 
\hline \\ 
Average pool. & 90.80\% & 90.81\% & 90.80\% & 90.80\%\\ 
Attention & \textbf{92.15\%} & \textbf{92.16\%} & \textbf{92.15\%}  & \textbf{92.15\%}\\
\hline \\
\end{tabular} 
\end{minipage}
\caption{Weighted average classification scores for temporal Average pooling and Attention method on speaker counting.}
\vspace{-0.7cm}
\label{tab:res}
\end{table}

\begin{figure*}
\vspace{-0.3cm}
\centering
\begin{tabular}{cc}
\hspace{-0.8cm}
\includegraphics[height=5.3cm,width=9cm]{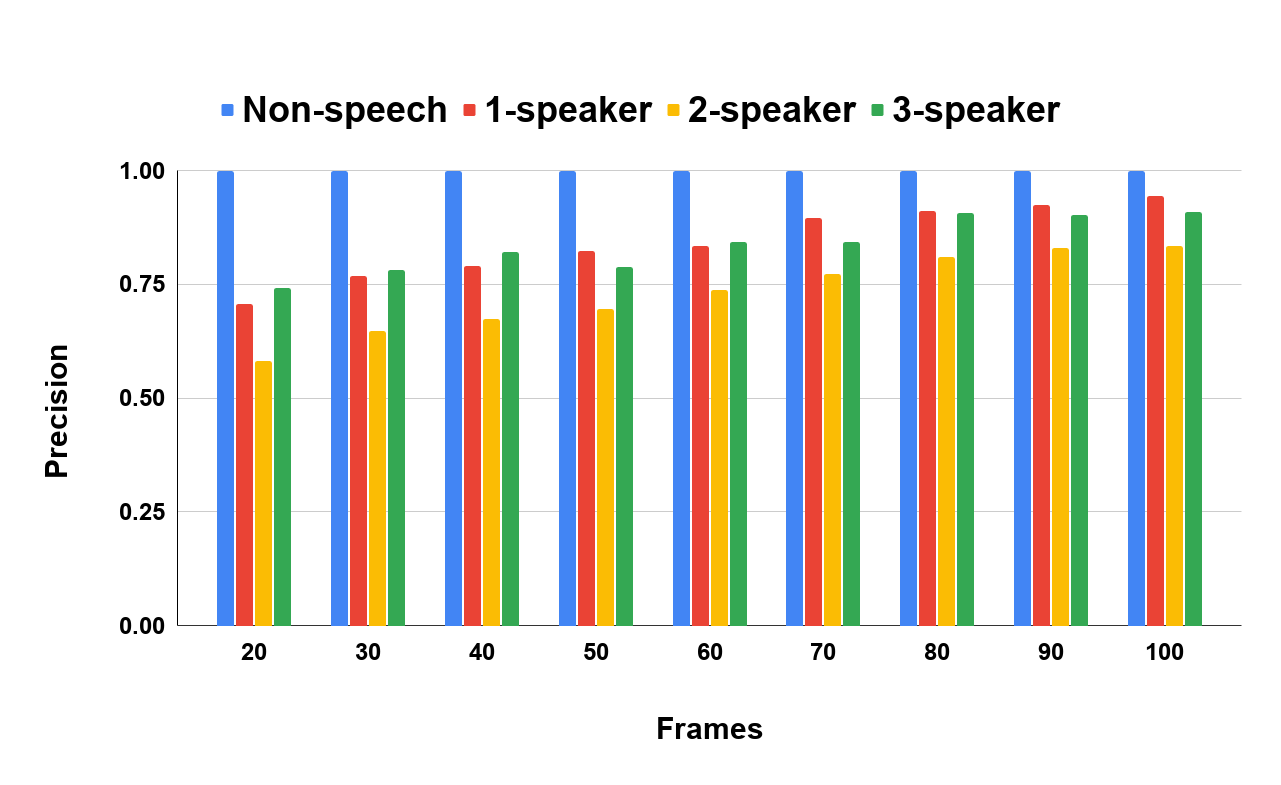}&
\hspace{-0.6cm}
\includegraphics[height=5cm,width=9cm]{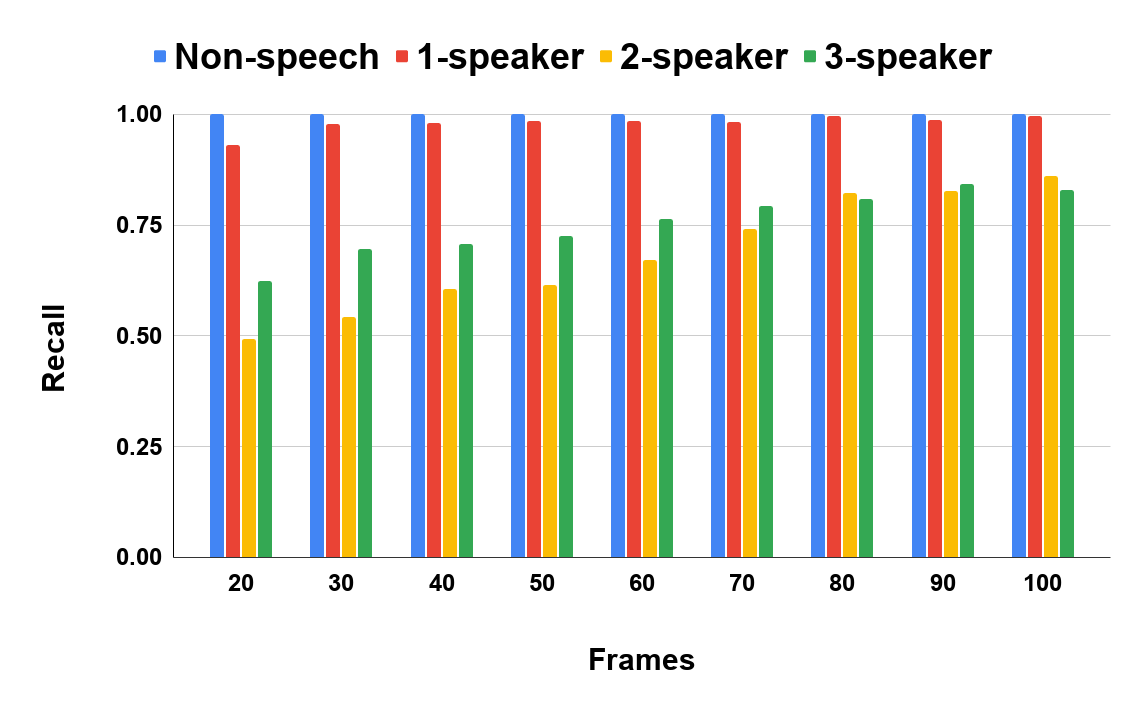} 
\hspace{-0.8cm}
\end{tabular}
\vspace{-0.6cm}
\label{figur}\caption{ The Precision and Recall for individual classes per different number of input frames using attention-guided CNN model.}
\vspace{-0.4cm}
\label{fig:att}
\end{figure*}

Next, the effect of attention mechanism is evaluated for speaker counting. As inferred from Table \ref{tab:res}, attention mechanism boosts speaker count estimation by almost 2-3\% absolute across all classification metrics for both real-time (20 frames) and offline (100 frames) scenarios. The Attention-guided feature vector is a condensed version of the feature map combining important information across the speech segment. Hence, the attention-guided network is able to achieve 76.15\% Weighted Accuracy and average Recall, 75.80\%  average Precision,  and 75.97\% F-score. All the classification metrics exceed 92\% for the attention-guided model in offline scenarios. 

\begin{figure}
\vspace{-0.4cm}
\centering
\includegraphics[width=8cm, height=5cm]{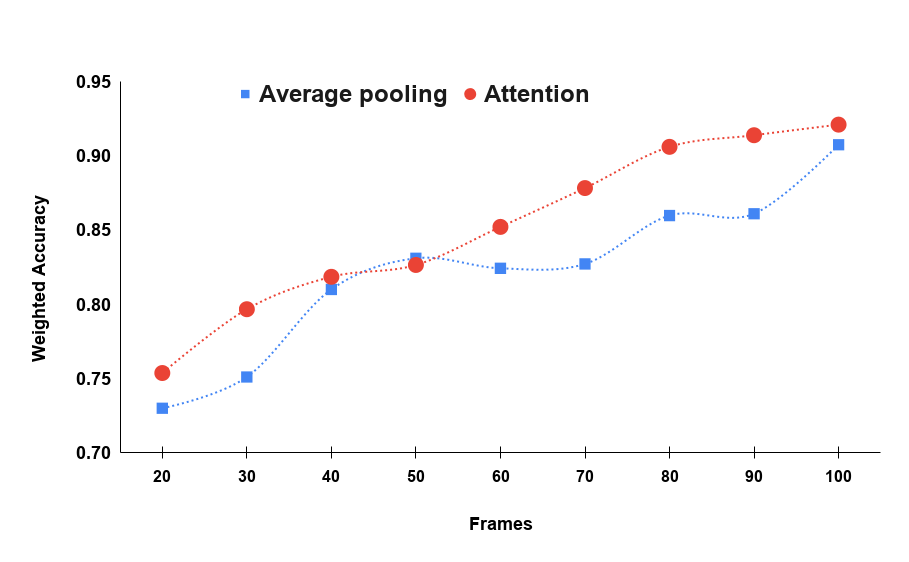}
\vspace{-0.9cm}
\caption{Comparing Weighted Accuracy per input frames for temporal Average pooling and Attention mechanism. }
\vspace{-0.5cm}
\label{fig:wacc-frame}
\end{figure}

The weighted accuracy for Average pooling over time (baseline) and Attention is plotted for varying input speech duration in Fig \ref{fig:wacc-frame}. Attention-guided CNN outperforms the Average pooling method for every duration of the input except for 50-frame speech, which achieves the same performance as the baseline. Since the chosen number of consecutive frames from the recording is random, the test set of 50-frame input speech might be probably selected from a rather easy portion of the dataset, therefore the superiority of the attention mechanism may not be fully revealed. However, for the challenging task of real-time speaker counting (20-30 frames), Attention consistently outperforms basic temporal Average pooling.  Fig \ref{fig:att} depicts performance of attention-guided CNN for individual classes in the real-time scenario (20-frame). According to this figure, Precision and Recall for non-speech achieves a perfect score, confirming the validity of the proposed solution also achieves effective Speech Activity Detector (SAD). Precision and Recall of the  2-speaker class is the lowest, because those segments are potentially confused with the 3-speaker class, especially if the speakers have the same gender and age. However, both Precision and Recall for the 2-speaker class exceeds 50\% which is twice the rate for random guess for a 4-class classification task. Also, the model is able to retrieve most single speaker segments with a Recall score close to 1. As expected, both Precision and Recall improves among all classes by increasing test duration (i.e., number of input frames). In conclusion, the experiments demonstrate the effectiveness of Attention mechanism in estimating the active speaker count for various speech duration in a multi-speaker scenarios.

Since each speaker count estimation algorithm in the literature is designed for a particular application, they usually consider specific assumptions to maximize different criteria. For example, the unsupervised Crowd++ \cite{xu2013crowd++} was an Android-based solution implemented in smartphones and tablet computers, and was tested on data collected from those devices. Crowd++ was designed to address the specific challenges faced in smartphones such as phone's location e.g., in or out of a pocket or bag, and the distance between the speakers and smartphones. In Crowd++, the authors attempted to find a rough estimate of the number of speakers in non-overlap speech recordings, and they used Error Count Distance (ECD) to evaluate their method. They showed that Crowed++ can estimate the speaker count with ECD of 1 manifesting that the estimated speaker count is usually one speaker less or more than the actual number of speakers. Since our proposed Attention-guided CNN is designed to estimates the exact number of co-current active speakers in overlap speech as a pre-processing step for speech separation and other speech technology systems, in our context, ECD of 1 is considered as a false and unreliable speaker count estimate. Also, CountNet \cite{stoter2018countnet} is a probabilistic model that uses Maximum A Posteriori (MAP) to estimate the number of active speakers. Since CountNet formulates the source count estimation as a regression problem, the authors used Mean Absolute Error (MAE) for evaluating the performance. Their goal was to find the closest estimate with the highest probability for the number of co-current speakers. However, as mentioned previously, in our application, we formulate this problem as a classification task, and maximize CrossEntropyLoss to find the exact number of speakers. Therefore, no matter how close the output estimate is to the actual number of speakers, if it is not exactly the same, we consider the estimated speaker count as an incorrect estimation, and those examples are reflected in the backpropagation process to update the network parameters for optimizing the model to estimate the precise co-current speaker count. 

\section{Conclusion}
\label{sec:con}
In this study, we proposed an attention-guided Convolutional Neural Network for real-time, single-channel, speaker count estimation in multi-speaker cocktail-party scenario. The proposed system includes three components: first a CNN model extracts higher-level information from spectral features of speech. Second, a temporal aggregation block is modeled using an attention mechanism, and third, a non-linear classifier was formulated using a fully connected network. All three elements of the proposed architecture are trained end-to-end by optimizing the log-likelihood loss of the classes. The proposed attention-guided method achieves almost 3\% absolute improvement over the conventional average pooling on segments as short as 20 frames (200ms) with 76.15\% Weighted Accuracy and average Recall, 75.80\% Precision, and 75.97\% F-score. 
Our proposed solution achieves 92\% across all classification metrics on speech segments longer than 1 second.

\bibliographystyle{IEEEtran}
\bibliography{mybib}
\end{document}